\documentclass[a4paper]{jpconf}
\usepackage{graphicx}
\usepackage{amsmath,amssymb}
\newcommand{\bra}[1]{\langle {#1} |}
\newcommand{\ket}[1]{| {#1} \rangle}

\def\vect#1{{\mbox{\boldmath $#1$}}}

\begin{document}
\title{Study of pygmy dipole resonance \\
with a new time-dependent mean field theory}

\author{Shuichiro Ebata$^{1,2}$,
Takashi Nakatsukasa$^{2,3}$
and Tsunenori Inakura$^{2}$}

\address{$^1$ Center for Nuclear Study, University of Tokyo, Wako-shi 351-0198, Japan}
\address{$^2$ RIKEN Nishina Center, Wako-shi 351-0198, Japan}
\address{$^3$ Center for Computational Sciences, University of Tsukuba,
Tsukuba 305-8571, Japan}

\ead{ebata@cns.s.u-tokyo.ac.jp}

\begin{abstract}
We carry out a systematic investigation on the low-energy electric dipole
strength, which is often called pygmy dipole resonances (PDR),
using the canonical-basis time-dependent Hartree-Fock-Bogoliubov (Cb-TDHFB)
method.
The Cb-TDHFB is a new method which is derived from TDHFB with
an approximation analogous to the BCS theory
that the pair potential is assumed to be diagonal in the time-dependent 
canonical basis. We apply the method to linear-response calculation for even-even nuclei.
We report the neutron-number dependence of PDR in 
light ($A < 70$) and heavy isotopes ($A > 100$) around $N=82$.
\end{abstract}
\section{Introduction}

Recent progress in radioactive isotope facilities and
experimental techniques allows us to investigate properties of new
exotic isotopes.
In the neutron-rich nuclei, the neutron-skin and halo structures are
known to be developed \cite{Tani85}. 
According to those exotic structures in the ground state,
new types of elementary modes of excitation are expected.
The low-lying electric dipole ($E1$) modes,
which are often called the pygmy dipole resonances (PDR),
have been extensively studied as one of such characteristic modes
of excitation in exotic nuclei \cite{Naka06}. 
However, the structure and the emergence mechanism of the PDR are still
elusive at present.
In this article, we perform a systematic investigation on the PDR
to elucidate its characteristic features.

In order to analyze and predict excited states in unstable nuclei, 
we use a time-dependent mean-field approach.
The time-dependent Hartree-Fock (TDHF) theory
has been extensively utilized to
study nuclear many-body dynamics \cite{Neg82}.
Recently it has been revisited with modern energy density functionals
and more accurate description of nuclear properties
has been achieved \cite{NY05,Mar05,UO06,UO07,WL08,CC09}.
However, it neglects the pairing interactions in particle-particle and hole-hole channels,
which are important for describing properties of open-shell heavy nuclei. 
It is well-known that the time-dependent Hartree-Fock-Bogoliubov (TDHFB) 
theory \cite{BR86} properly takes into account the pairing correlation.
Unfortunately, the practical calculation with the TDHFB are very limited \cite{HN07,ASC08}, 
because it requires us to deal with the time evolution of ``infinite'' number
of quasi-particle orbitals, in principle.
In practice, the number of the quasi-particle orbitals is approximately
equal to the dimension of the adopted one-body model space.

Recently, we have developed the canonical-basis TDHFB (Cb-TDHFB) theory using a simple approximation 
to avoid to treat the ``infinite'' orbitals \cite{Eba10}. 
The Cb-TDHFB is a time-dependent version of the BCS approximation \cite{RS80} for
the Hartree-Fock-Bogoliubov theory. 
Namely, we neglect off-diagonal elements of the pair potential in the
time-dependent canonical-basis. 
In this paper,
we use the Cb-TDHFB method to perform the linear-response calculation
with
the full Skyrme functional of the SkM* parameter set.
We report the neutron number dependence of the low-lying $E1$ strength 
for light and heavy neutron-rich isotopes. 

The paper is organized as follows.
In Sec.~\ref{aba:sec2}, we present the basic equations of the Cb-TDHFB 
method, their conservation lows and the numerical procedure of the present work.
In Sec.~\ref{sec:numerical_results},
we present numerical results of the real-time calculations of the
$E1$ mode for many isotopes.
Finally, the summary is given in Sec.~\ref{sec:summary}.

\section{Cb-TDHFB equations}\label{aba:sec2}
\subsection{Basic equations}

The Cb-TDHFB can be derived from the TDHFB equation on 
the assumption that the pair potential can be approximated to be diagonal in the canonical-basis.
The readers should be referred to Ref. \cite{Eba10} for more details of the derivation.

The TDHFB equation can be written in terms of the generalized density
matrix $R(t)$ and the HFB Hamiltonian ${\cal H}(t)$ as follows \cite{BR86}:
\begin{equation}
i\frac{\partial}{\partial t} R(t) = \left[ {\cal H}(t), R(t) \right] .
\end{equation}
This is equivalent to the following equations for
one-body density matrix $\rho(t)$ 
and the pairing-tensor matrix $\kappa(t)$.
\begin{subequations}
\begin{eqnarray}
\label{TDHFB_1}
i\frac{\partial}{\partial t}\rho(t) &=&
  [h(t),\rho(t)] + \kappa(t) \Delta^*(t) - \Delta(t) \kappa^*(t) ,\\
\label{TDHFB_2}
i\frac{\partial}{\partial t}\kappa(t) &=&
 h(t)\kappa(t)+\kappa(t) h^*(t) + \Delta(t) (1-\rho^*(t)) - \rho(t) \Delta(t) .
\end{eqnarray}
\end{subequations}
Here, $h(t)$ and $\Delta(t)$ are the single-particle Hamiltonian and
the pair potential, respectively. 
Introducing the time-dependent canonical states $\ket{\phi_k(t)}$ and
$\ket{\phi_{\bar k}(t)}$,
we express the TDHFB state in the canonical (BCS) form as
\begin{equation}
\ket{\Psi(t)}=\prod_{k>0} \left\{
u_k(t) + v_k(t) c_k^\dagger(t) c_{\bar k}^\dagger(t) \right\} \ket{0} .
\end{equation}
Here, it should be noted that the pair of states, $k$ and $\bar k$,
are not necessarily relate to each other by the time-reversal,
$\ket{\phi_{\bar{k}}(t)}\neq T\ket{\phi_{k}(t)}$.
The diagonal approximation of the pair potential lead Eqs. (\ref{TDHFB_1})
and (\ref{TDHFB_2}) to the following equations:
\begin{subequations}
\label{Cb-TDHFB}
\begin{eqnarray}
\label{dphi_dt}
&&i\frac{\partial}{\partial t} \ket{\phi_k(t)} =
(h(t)-\eta_k(t))\ket{\phi_k(t)} , \quad\quad
i\frac{\partial}{\partial t} \ket{\phi_{\bar k}(t)} =
(h(t)-\eta_{\bar k}(t))\ket{\phi_{\bar k}(t)} , \\
\label{drho_dt}
&&
i\frac{d}{dt}\rho_k(t) =
\kappa_k(t) \Delta_k^{\ast}(t)
-\kappa_k^{\ast}(t) \Delta_k(t) , \\
\label{dkappa_dt}
&&
i\frac{d}{dt}\kappa_k(t) =
\left(
\eta_k(t)+\eta_{\bar k}(t)
\right) \kappa_k(t) +
\Delta_k(t) \left( 2\rho_k(t) -1 \right) .
\end{eqnarray}
\end{subequations} 
These basic equations determine the time evolution of
the canonical states, $\ket{\phi_k(t)}$ and $\ket{\phi_{\bar k}(t)}$,
their occupation $\rho_k(t)\!=\!|v_{k}(t)|^{2}$,
and pair probabilities $\kappa_k(t)\!=\!u_{k}(t)v_{k}(t)$.
The time-dependent gap $\Delta_k(t)$ is calculated as
\begin{equation}
\Delta_k(t)=-\sum_{l>0} \kappa_l(t)
{\bar v}_{k{\bar k},l{\bar l}} ,
\end{equation}
using the anti-symmetrized two-body interaction $\bar v$ \cite{Eba10}.
The real function of time $\eta_k(t)$ is defined as
\begin{equation}
\label{phase_eta}
\eta_k(t)=\bra{\phi_k(t)} h(t) \ket{\phi_k(t)} + i \left\langle \frac{\partial \phi_{k}(t)}{\partial t} \Big| \phi_{k}(t) \right\rangle.
\end{equation}
This is an arbitrary gauge parameter and
depends on the phase of the canonical states.
Note that the symbols, $(\rho, \kappa, \Delta)$, indicate
matrices with two indices in Eqs. (\ref{TDHFB_1}) and (\ref{TDHFB_2}),
but vectors with a single index in Eqs. (\ref{dphi_dt}), (\ref{drho_dt}),
and (\ref{dkappa_dt}).
Similar equations can be found 
in Ref.~\cite{BF76} for a very schematic pairing energy functional.

\subsection{Pairing energy functional and cut-off function}
\label{sec:procedure}

We adopt a Skyrme functional with the SkM* parameter set for the
particle-hole channels.
For the pairing energy functional, we adopt a simple functional 
of a form  
\begin{equation}
\label{E_G}
E_{\rm pair}(t)=-\sum_{k,l>0} G_{kl} \kappa_k^*(t) \kappa_l(t) 
=-\sum_{k>0} \kappa_k^*(t) \Delta_k(t) ,
\quad
\Delta_k(t)= \sum_{l>0} G_{kl} \kappa_l(t), 
\end{equation}
with a following special gauge condition \cite{Eba10},
\begin{equation}
\label{gauge_fix}
\eta_k(t)=\varepsilon_k(t)=\bra{\phi_k(t)} h(t) \ket{\phi_k(t)}, \quad
\eta_{\bar k}(t)=\varepsilon_{\bar k}(t) 
=\bra{\phi_{\bar k}(t)} h(t) \ket{\phi_{\bar k}(t)}. 
\end{equation} 
Here, $G_{kl}=G f(\varepsilon_k^0) f(\varepsilon_l^0)$ and the
constant $G$ is determined by the smoothed pairing method \cite{Eba10}.
The cut-off function $f(\varepsilon_k^0)$ depends on the single-particle energy
of the canonical state $k$ at the HF+BCS ground state. The cut-off function $f(\varepsilon)$ is written as 
\begin{equation}
f(\varepsilon)=\left( 1+\exp\left[ 
\frac{\varepsilon - \epsilon_{\rm c}}{0.5 \mbox{ MeV}} \right]\
 \right)^{-1/2} \theta (e_{\rm c}-\varepsilon),
\end{equation}
with the cut-off energies 
\begin{eqnarray}
\epsilon_{\rm c} = \tilde\lambda+5.0\ {\rm MeV},\quad
e_{\rm c} = \epsilon_{\rm c}+2.3\ {\rm MeV} ,
\end{eqnarray}
where $\tilde\lambda$ is the average of the highest occupied level
and the lowest unoccupied level in the HF state.
Here, the cut-off parameter $e_{\rm c}$ is necessary to prevent
occupation of spatially unlocalized single-particle states,
known as the problem of unphysical gas near the drip line.
For neutrons, if $e_{\rm c}$ becomes positive, we replace it by zero \cite{TTO96}.

\subsection{Properties of the Cb-TDHFB equations}
\label{sec:properties}

The Cb-TDHFB equations possess the following properties \cite{Eba10}:
\begin{enumerate}
\item Conservation law
 \begin{enumerate}
 \item Conservation of orthonormal property of the canonical states
 \item Conservation of average particle number
 \item Conservation of average total energy
 \end{enumerate}
\item The stationary solution corresponds to the HF+BCS solution.
\item In the limit of $\Delta=0$, they are equivalent to TDHF.
\item In the small-amplitude limit, 
          if the ground state is in the normal phase, the equations are
          identical to the particle-hole, particle-particle, and hole-hole
          RPA with the BCS approximation.
\end{enumerate}

\subsection{Numerical details}
For numerical calculations, we extended the computational program of the
TDHF in the three-dimensional (3D) coordinate-space representation \cite{NY05}
to include the BCS-type pairing correlations.
The ground state is first constructed by the HF+BCS calculation.
Then, we add a weak impulse isovector dipole field $V_{E1}(t)$ to 
the ground state. $V_{E1}(t)$ is written as 
\begin{equation}
\label{V_E1}
V_{E1}(t) = -\eta F_{E1} \delta(t), \hspace{5mm}
F_{E1}=
\begin{cases}
  (Ne/A) r_i  & \text{for protons} \\
- (Ze/A) r_i  & \text{for neutrons}
\end{cases} ,
\end{equation}
where $i=(x,y,z)$. 
We solve the Cb-TDHFB equations in real time and real space, then calculate the linear-response of the nucleus. 
We calculate the time evolution of the expectation value of $F_{E1}$ 
under the external field $V_{E1}(t)$ with $\eta\ll1$ and obtain the strength function $S(E1;E)$ by Fourier transformation 
using an exponential smoothing with $\Gamma=1$ MeV: $V_{E1}(t) \rightarrow V_{E1}(t)e^{-\Gamma t/2}$. 

In this work, we use the 3D Cartesian coordinate for the canonical states, 
$\phi_{l}(\vect{r},\sigma;t)=\bra{\vect{r},\sigma}\phi_{l}(t)\rangle$ with $\sigma=\pm 1/2$. 
For light isotopes ($A < 70$), 
the coordinate space is discretized in the square mesh of $\Delta x=\Delta y=\Delta z=0.8$ fm 
in a sphere with radius of 12 fm. 
For heavy isotopes ($A > 100$), 
we choose the square mesh of $\Delta x=\Delta y=\Delta z=1.0$ fm in a sphere with radius of 15 fm.

\section{Neutron-number dependence of the low-lying $E1$ strength}
\label{sec:numerical_results}

In this section, we discuss a systematic trend of the low-lying $E1$ strength
as a function of neutron number.
To quantify the strength of PDR,
the following ratio of the low-lying $E1$ strength is used:
\begin{eqnarray}
\label{Ratio}
 \frac{m_{1}(E_{\rm c})}{m_{1}} \equiv \frac{\int^{E_{\rm c}} E\times S(E1; E) dE}{\int E\times S(E1; E) dE} \times 100 \ [\%],
\end{eqnarray}
where $E_{\rm c}$ is a cut-off energy. We take $E_{\rm c}=$10 MeV for all calculations. \\[-1cm]
\begin{figure}[h]
 \begin{minipage}{0.45\hsize}
  \begin{center}
   \includegraphics[keepaspectratio,width=55mm, angle=-90]{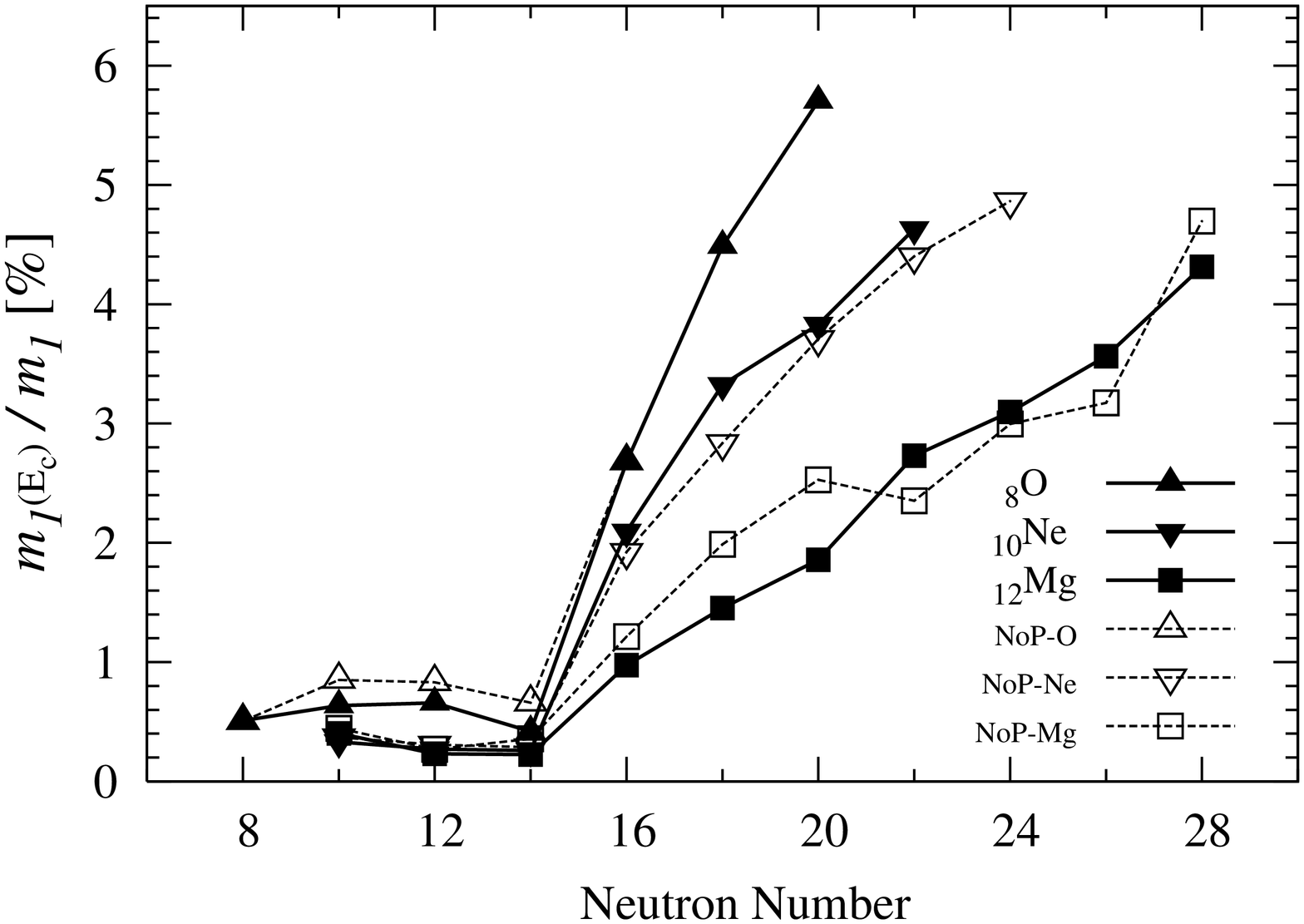}
   \caption{\footnotesize Neutron number dependence of $m_{1}(E_{\rm c})/m_{1}$ defined in Eq.\eqref{Ratio}
for O, Ne and Mg isotopes}
\label{fig:O-Mg_Pyg}
   \end{center}
 \end{minipage}\hspace{5mm}
 \begin{minipage}{0.45\hsize}
\ \\[-15mm]
   \begin{center}
\includegraphics[keepaspectratio,width=55mm, angle=-90]{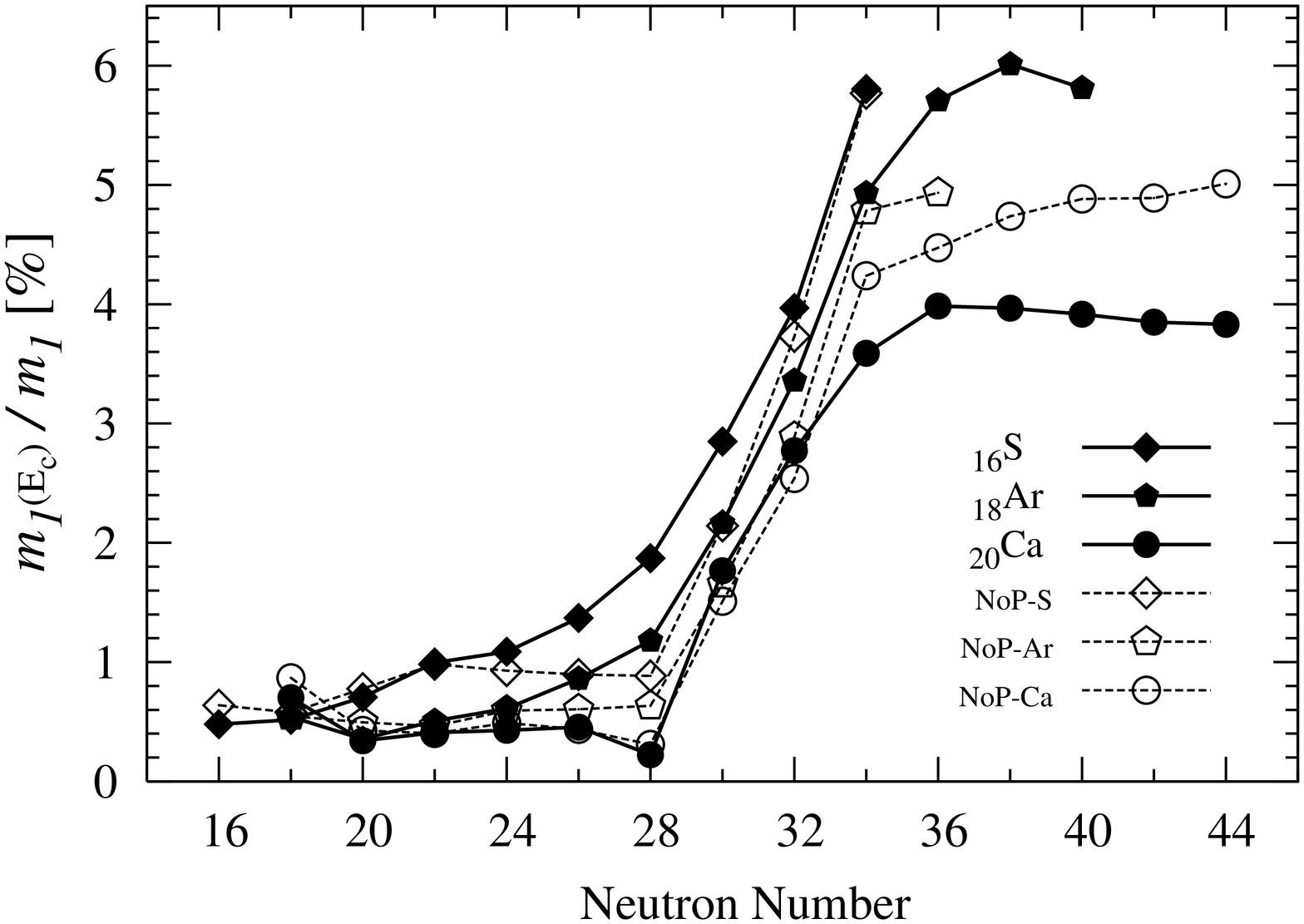}
   \caption{\footnotesize Same as Fig. 1 but for S, Ar and Ca isotopes.}
\label{fig:S-Ca_Pyg}
  \end{center}
 \end{minipage}
\end{figure} \\[-1cm]

\subsection{Light to medium nuclei $(A < 70)$}

For light nuclei with $Z\leq 40$, the systematic analysis has been carried
out in Ref. \cite{Ina11}.
The calculation was based on the finite amplitude method
\cite{Nak07,Ina09} with the Skyrme energy functional
of SkM*, however, the pairing correlation was neglected.
Thus, it would be desirable to investigate the effect of the pairing
correlations on the low-energy $E1$ strength.

Figure 1 shows the results for O, Ne and Mg isotopes as a function of neutron number and 
Fig. 2 shows those for S, Ar and Ca isotopes. 
Solid lines with filled symbols indicate the results of Cb-TDHFB calculation and 
dashed lines with open symbols are those of Hartree-Fock plus random phase approximation (HF+RPA) 
without the pairing correlation \cite{Ina11}. 
In both cases,
we see that the nuclei with $N=$8 - 14 have the fractions of low-lying $E1$ strength less than $1.0\%$, 
and a sudden jump of the fraction at $N=14\rightarrow 16$ on each isotopic chain in Fig. 1. 
The neutron number $N=16$ corresponds to the occupation of the $s_{1/2}$ orbit.
Figure 1 shows a neutron shell effect on low-lying $E1$ strength. 
This suggests that the position of the Fermi level plays an
important role for the emergence of low-lying $E1$ strength. 

A similar behavior is also seen in Fig. 2, however, is slightly changed by the
pairing correlations.
Namely, the nuclei with $N=$16 - 28 have small fractions in HF+RPA, 
and the sudden jump from $N=28$ to $30$ is clearly observed.
In Ca isotopes, the fraction has almost constant values at $N\geq$ 36. 
The neutron number $N=30$ corresponds to the occupation of the $p_{3/2}$ orbit, 
and $N=36$ to the $f_{5/2}$ orbit. 
The occupation of the orbits with small orbital angular momentum
is important for the emergence of low-lying $E1$ strength. 
 
The effects of the pairing correlation is mostly seen around $N=28$ in Fig. 2.
The low-lying $E1$ strength increases smoothly around $N=28$
in S and Ar isotopes in the Cb-TDHFB calculation,
while we see sudden increases from $N=28$ to $30$ in the calculation without
the pairing (HF+RPA).
This smooth evolution of the low-lying strength is caused by the fractional
occupation probabilities 
of the single-particle orbits, due to the pairing correlation. 
For lighter nuclei in Fig. 1,
the pairing effect is seen in Ne and Mg isotopes.
This is mostly due to different shapes in the ground state. 
Actually, for $N=16-20$,
the calculated ground states in the HF+BCS are spherical,
but those in the HF are deformed. 
\begin{figure}[t]
\centerline{
\includegraphics[width=7.6cm,angle=-90]{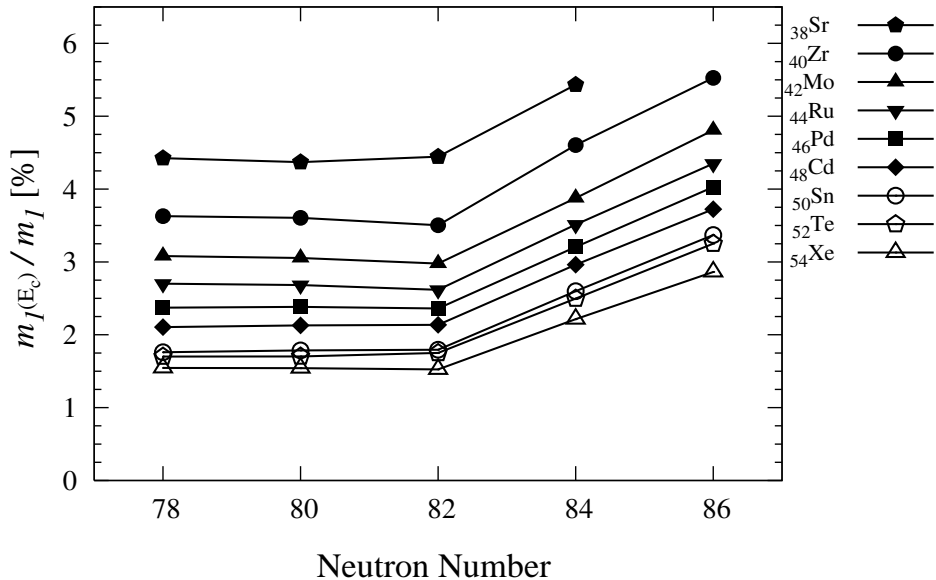}
}
\caption{\label{fig:Heavy_Pyg}
Same as Fig. 1 but for Sr, Zr, Mo, Ru, Pd, Cd, Sn, Te and Xe isotopes ($N=$ 78 - 86) 
are shown as a function of neutron number. All results are obtained by using Cb-TDHFB.}
\end{figure}

\subsection{Heavy nuclei $(A > 100)$}

Figure 3 shows the results for Sr, Zr, Mo, Ru, Pd, Cd, Sn, Te and Xe isotopes with $N=$ 78 - 86, 
calculated with Cb-TDHFB. 
Again,
we can clearly see a sudden increase of the PDR ratio from $N=$82 to 84,
for all isotopic chains. 
The kink position corresponds to the well-known neutron magic number $N=82$.
Even in the case of heavy isotopes, the neutron shell effects play a
very important role in the low-lying $E1$ strength. 
However,
there are constant but significant PDR fractions for $N=$78 - 82. 
Although $^{132-136}$Xe are stable nuclei,
they have $1.5\%$ of low-lying $E1$ strength. 
This fraction even increases as the proton number decreases.
We suppose that these constant fractions have a different origin from
the one observed in the light neutron-rich nuclei. 
This is currently under investigation.

\section{Summary} 
\label{sec:summary}

We presented an approximate and feasible approach to the TDHFB;
the canonical-basis TDHFB method.
We calculated the $E1$ strength distribution in neutron-rich isotopes of light and heavy nuclei, 
using the real-time real-space method.
The low-lying $E1$ strength has a characteristic neutron number dependence. 
In light neutron-rich isotopes with $Z=$ 8 - 20, the fraction of the
energy-weighted $E1$ strength below 10 MeV increases at $N=16$ in O, Ne and Mg isotopes and at $N=30$ in S, Ar and Ca isotopes. 
These numbers indicate that the occupation of the single-particle states with
small orbital angular momentum  is very important for the emergence 
of the low-lying $E1$ strength. 
The heavy nuclei with neutron excess around $N=82$ have been also
investigated.
There is a similar jump in the PDR fraction at $N=82\rightarrow 84$.
However, a significant amount of the PDR fraction is also observed in
nuclei before the jump with $N\leq 82$.

\ack
This work is supported by High Performance Computing Infrastructure 
Strategic Program Field 5,
and by Grant-in-Aid for Scientific Research(B)
(No. 21340073) and on Innovative Areas (No. 20105003).
The computational resources were provided by the RIKEN Integrated Cluster
of Clusters (RICC) and by the Joint Research Program
at Center for Computational Sciences, University of Tsukuba.

\section*{References}
\bibliographystyle{iopart-num}
\bibliography{nuclear_physics,myself}

\end{document}